\documentclass[sigconf,natbib=true]{acmart}

\usepackage{caption}
\usepackage{amsmath}
\usepackage{multirow}
\usepackage{breqn}
\usepackage{graphicx}  
\usepackage{float}

\newlength{\beforesecskip}
\newlength{\aftersecskip}
\usepackage{color}
\usepackage{bbm}
\usepackage{multirow}
\usepackage[inline]{enumitem}
\usepackage{graphicx}
\usepackage{sistyle}
\usepackage[ruled]{algorithm2e}
\usepackage{booktabs}
\usepackage{caption}
\SIthousandsep{,}
\usepackage{makecell}
\usepackage[skip=0pt]{caption}
\usepackage{amsmath}
\usepackage{hyperref}
\usepackage{array} 
\usepackage{graphicx}
\usepackage{acronym}
\usepackage[export]{adjustbox}

\AtBeginDocument{%
  \providecommand\BibTeX{{%
    \normalfont B\kern-0.5em{\scshape i\kern-0.25em b}\kern-0.8em\TeX}}}

\makeatletter
\g@addto@macro\normalsize{%
  \abovedisplayskip 2pt plus1pt 
  \belowdisplayskip 2pt plus1pt
  \abovedisplayshortskip  2pt plus1pt%
  \belowdisplayshortskip  1pt plus1pt
}
\makeatother

\AtBeginDocument{%
  \providecommand\BibTeX{{%
    Bib\TeX}}}

\setcopyright{acmlicensed}
\copyrightyear{2018}
\acmYear{2018}
\acmDOI{XXXXXXX.XXXXXXX}

\acmConference[Conference acronym 'XX]{Make sure to enter the correct
  conference title from your rights confirmation email}{June 03--05,
  2018}{Woodstock, NY}
\acmISBN{978-1-4503-XXXX-X/18/06}


\author{Da Li}
\affiliation{
  \institution{CAS Key Lab of Network Data Science and Technology, ICT, CAS}
  \institution{University of Chinese Academy of Sciences}
  \city{Beijing}
  \country{China}
}
\email{lida.ucas@gmail.com}

\author{Keping Bi}
\orcid{}
\affiliation{
	\institution{CAS Key Lab of Network Data Science and Technology, ICT, CAS}
	\institution{University of Chinese Academy of Sciences}
	\city{Beijing}
	\country{China}
}
\email{bikeping@ict.ac.cn}

\author{Jiafeng Guo}
\orcid{}
\affiliation{
	\institution{CAS Key Lab of Network Data Science and Technology, ICT, CAS}
	\institution{University of Chinese Academy of Sciences}
	\city{Beijing}
	\country{China}
}
\email{guojiafeng@ict.ac.cn}

\author{Xueqi Cheng}
\orcid{0000-0002-5201-8195}
\affiliation{
	\institution{CAS Key Lab of Network Data Science and Technology, ICT, CAS}
	\institution{University of Chinese Academy of Sciences}
	\city{Beijing}
	\country{China}
}
\email{cxq@ict.ac.cn}

\begin{document}

\title{Bridging Queries and Tables through Entities in Table Retrieval}
\begin{abstract}
\textbf{Table retrieval} is essential for accessing information stored in structured tabular formats; however, it remains less explored than text retrieval. The content of the table primarily consists of phrases and words, which include a large number of entities, such as time, locations, persons, and organizations. Entities are well-studied in the context of text retrieval, but there is a noticeable lack of research on their applications in table retrieval.
In this work, we explore how to leverage entities in tables to improve retrieval performance. First, we investigate the important role of entities in table retrieval from a statistical perspective and propose an entity-enhanced training framework. Subsequently, we use the type of entities to highlight entities instead of introducing an external knowledge base. Moreover, we design an interaction paradigm based on entity representations. Our proposed framework is plug-and-play and flexible, making it easy to integrate into existing table retriever training processes.
Empirical results on two table retrieval benchmarks, NQ-TABLES and OTT-QA, show that our proposed framework is both simple and effective in enhancing existing retrievers. We also conduct extensive analyses to confirm the efficacy of different components. Overall, our work provides a promising direction for elevating table retrieval, enlightening future research in this area.

\end{abstract}


\maketitle

\section{Introduction}
Table retrieval is essential for accessing the extensive information stored in tabular formats, playing a key role in numerous daily applications~\citep{cafarella2008webtables,jauhar2016tables,zhang2020web}. The Natural Questions dataset shows that 25.6\% of information requests require tables for complete answers~\citep{kwiatkowski-etal-2019-natural}. The relevant table serves as the foundation for table-related tasks such as Table Fact Verification~\citep{chen2020tabfact} and TableQA~\citep{cafarella2008webtables,jauhar2016tables} that even need to translate queries into SQL. 
However, table retrieval is underexplored compared to text retrieval. In the era of Large Language Models (LLMs), TableGPT2~\citep{su2024tablegpt2largemultimodalmodel} and TableLLM~\citep{zhang2024tablellmenablingtabulardata} are designed to handle complex tables to address users’ information need. The ability to retrieve query-relevant tables from a large corpus is a crucial factor in determining whether these LLMs can operate in an end-to-end manner to serve users effectively.

There are prior works exploring table retrieval. Existing approaches can be categorized into two main technical routes: one emphasizes the refinement of encoding strategies, while the other aims to improve training strategies. The former attempts to distinguish table rows and columns by utilizing specialized network layers such as TAPAS~\citep{Herzig_2020}, which incorporates row and column embeddings to effectively capture tabular structures. The latter improves performance by modifying the training patterns. \citet{herzig2021open} and \citet{chen2023bridge} attempt to optimize table retrievers by special pre-training tasks for the retrieval and tables. In addition to single-vector retrievers, SSDR~\citep{jin2023enhancing} represents queries and tables as multi-vectors to strengthen fine-grained matching. With the increasing complexity of queries, a single table often fails to fully satisfy user information needs, prompting the introduction of multi-table retrieval as a new task~\citep{chen2024tableretrievalsolvedproblem}. Recently, the task of table reranking has attracted considerable attention. Some studies assert that they optimize table retrieval, but in fact, they focus on table reranking, which has caused some confusion. But some of them can be adapted for table retrieval. TaBERT~\citep{yin2020tabert} and StruBERT~\citep{Trabelsi_2022} partition tables into rows and columns, and then aggregate the row and column representations using a specialized self-attention mechanism. Considering the layout and structure of tables, GTR~\citep{Wang_2021} employs graphs to represent their row and column relationships.

The table content consists mainly of phrases and words, which include a large number of entities. Tables are designed to represent collections of similar data, facilitating the grouping of entities of the same type, such as time, place, and organization. Entities are widely utilized in text retrieval, resulting in significant performance improvement~\citep{gonccalves2018improving, DBLP:journals/corr/XiongCL17}, while little work has been done to examine the impact of entities on table retrieval. 
Before pre-trained models (PLMs) were widely used, Table2Vec~\citep{Zhang_2019} represents tables as sequences of entities, enhancing retrieval performance by learning entity representations, the integration of entities and PLMs remains underexplored. We investigated the potential value of entities for table retrieval through preliminary studies from a statistical perspective. This provides insights into ways to enhance table retrieval by leveraging entities.

In this work, our goal is to improve the performance of table retrieval by leveraging entities. This presents several challenges. First, how can we effectively leverage entities in tables collected automatically from the web? Second, in what ways can entities contribute to improving retriever performance?
built upon our statistical research, we propose an entity-enhanced table retrieval training framework to address these challenges. This framework has two main components. First, we design an entity type embedding to inject type information into entities appearing in queries and tables. Additionally, we propose an entity-based interaction paradigm based on the different backbones of retrievers to highlight the representation of entity information within the final representations of queries and tables. In the inference stage, only the entity type embedding is preserved, which guarantees efficiency for online retrieval. 
Empirical results indicate that the proposed framework effectively enhances the existing retrievers with entity-matching capabilities, significantly improving retrieval performance. 
Additionally, its flexible, plug-and-play characteristics allow seamless integration with various backbones of table retrievers. Our main contributions are summarized as follows:
\begin{itemize}[leftmargin=*]
    \item We conducted a statistical analysis to investigate the distribution of entities in tables and their effectiveness in distinguishing between relevant and irrelevant retrieval tables.
    \item We propose a novel entity-enhanced training framework that significantly improves the performance of existing table retrievers.
    \item Extensive experiments show that our framework is effective when integrating with different types of backbone retrievers; we also conducted extensive analyses to show how this framework takes effect.
\end{itemize}

\section{Related Work} 
In this section, we review and summarize relevant advances in the areas covered by our methodology: table search and entity-enhanced retrieval.
\subsection{Table Search}
Table search, introduced by \citet{Zhang_2018}, refers to the process of identifying tables that are relevant to a user query within a large corpus of tables, aiming to satisfy the user’s information needs. Although much research and significant progress have been made in text search, the row and column structure of tables poses unique challenges, unlike plain text. The difficulty is in utilizing the structural characteristics of tables to enrich the representation to improve performance.

\subsubsection{Table Retrieval}
Table retrieval refers to the process of retrieving relevant tables for a specific query from a large-scale corpus of tables. Table retrievers independently encode queries and tables. Given the success of pre-trained language models (PLMs) in text retrieval, some methods have been adapted for table retrieval, achieving impressive performance. 

These adaptation methods are optimized for two aspects. On the one hand, some adaptations introduce additional neural network layers and representations to encode the structure of tables. TAPAS~\citep{Herzig_2020} uses row and column embeddings to encode table structure information and has competitive performance in table retrieval, categorization, and other table-related tasks. 
On the other hand, some methods narrow the gap between the model’s textual and tabular representations by adjusting the training data and methodologies. With specific data and training methods built for the tables, PLMs can automatically learn the structure of the input. Building on BERT, UTP~\citep{chen2023bridge} employs an unsupervised alignment loss to integrate tables and surrounding text, achieving a consistent cross-modal representation, DTR~\citep{herzig2021open} develops a table retriever based on TAPAS, utilizing the Inverse Cloze Task. For tables, which are complex two-dimensional structured data, single-vector representation focuses on global information, while multi-vector dense retrievers can represent fine-grained information. SSDR~\citep{jin2023enhancing} enhances fine-grained matching with multiple vectors leveraging syntactic and structural information of queries and tables. 
In addition to single-table retrieval, multi-table retrieval has also attracted attention from the community~\citep{chen2024tableretrievalsolvedproblem}.

\subsubsection{Table Reranking}
Recently, the task of table reranking has received increasing attention, and some of these studies could be applied to enhance table retrieval. TaBERT~\citep{yin2020tabert} encodes tables row by row, pairing the query with each row as input, and then aggregates the row representations using vertical self-attention. Building on TaBERT, StruBERT~\citep{Trabelsi_2022} takes a step further by slicing the table into columns and introducing a horizontal attention mechanism to aggregate information across columns. COTER~\citep{10.1145/3616855.3635796} introduces Conditional Optimal Transport to table reranking, simplifying table content and realizing performance improvements. 
Additionally, other approaches consider the table layout and structure, formulating tables as hypergraphs by defining various types of nodes and edges. GTR~\citep{Wang_2021} builds tabular graph structures to cover tabular information at different granularities based on the rows, columns, and cells of the table.
However, as they jointly encode the query and table, their design is tailored for table reranking tasks and can not be directly applied to table retrieval.

\subsection{Entity-Enhanced Retrieval}\label{related_entity_enhanced}
Entities refer to specific, recognizable objects with relatively independent context, and they enhance tasks such as question answering~\citep{yasunaga2021qa, zhang2022greaselmgraphreasoningenhanced}, recommendations~\citep{wang2023knowledge, Wang_2019}, and retrieval~\citep{gonccalves2018improving, DBLP:journals/corr/XiongCL17, Zhang_2019}.

In text retrieval, entities serve as an effective external feature to enhance retrieval performance. \citet{gonccalves2018improving} links the mentions in the query and document to corresponding entities in Wikipedia and combines bag-of-words (BoW) with bag-of-entities (BoE) representations for sparse retrieval. \citet{zhang2022greaselmgraphreasoningenhanced} improves the performance of sparse retrievers by matching and associating specific words in the text (e.g., locations, organizations, countries, etc.) with corresponding
entities in the knowledge base by entity linking. \citet{DBLP:journals/corr/XiongCL17} calculates the similarity between the query and the document based on the lexical terms and entities respectively, and finally fuses the similarity scores of these two parts in a weighted manner to generate the final relevance score. EQFE~\citep{10.1145/2600428.2609628}  links potential entities in a query to corresponding entities in the knowledge base. It extends the query by utilizing information about the entities in the linked knowledge base such as attributes, categories, associated entities, etc. \citet{tran2022dense} effectively improves the retrieval by fusing the text representation and the representation of different entity views. KGPR~\citep{fang2023kgpr} uses knowledge graphs (KGs) as additional inputs to provide the background knowledge. It uses an entity linking tool to identify entities in queries and passages and then utilizes subgraphs extracted from large KGs that are relevant to queries and passages to improve retrieval performance. Entities also have an important role in form retrieval. Table2Vec~\citep{Zhang_2019} attempts to transform tables into sequences of entities. Representations are constructed for the entities appearing in the table by the same approach as Word2Vec\citep{mikolov2013efficientestimationwordrepresentations}. 
While these methods effectively integrate external knowledge, retrieval performance remains dependent on the specific knowledge base. Our approach leverages entity-type information without requiring any additional knowledge base.

\section{Preliminary Study}\label{sec:prelim}
In this section, we analyze the characteristics of entities in table retrieval and text retrieval. These analyses provide valuable insights for the design of our framework. We utilized two widely used retrieval datasets for preliminary studies: MS MARCO~\citep{bajaj2018msmarcohumangenerated} for text and NQ-TABLES~\citep{herzig2021open} for table. They are collected from extensive web pages and search logs, with relevance annotations ensuring quality and usability. The statistical results can reflect the characteristics of entities in two different data formats under real-world scenarios. For both queries and corpus, we used spaCy~\citep{honnibal2017spacy} to identify potential entities within them. 

\subsection{Entity Coverage in Table Retrieval}
As a first step, we examine the frequency of entity occurrences in MS MARCO and NQ-TABLES.
The statistical results are shown in Table~\ref{table:table_entity_count}.

\begin{table}[htbp!]
\centering
\begin{adjustbox}{width=\linewidth}
\renewcommand{\arraystretch}{0.90}
\begin{tabular}{lrrrr} \toprule
                 & \multicolumn{2}{c}{MS MARCO} & \multicolumn{2}{c}{NQ-TABLES} \\ 
\cmidrule(lr){2-3} \cmidrule(lr){4-5} 
                 & Query        & Passage       & Query         & Table         \\ \midrule
Avg. \# Tokens   & 6.90         & 89.22         & 9.51          & 531.03        \\
Avg. \# Entities & 0.22         & 4.87          & 0.44          & 51.31         \\
Entity Coverage  & 26.63\%      & 51.11\%       & 42.96\%       & 84.36\%         \\ \bottomrule
\end{tabular}
\end{adjustbox}
 \caption{\label{table:table_entity_count}
Statistics of Entity in MS MARCO and NQ-TABLES.}
\end{table} 
The results reveal that tables are usually longer and contain more entities than passages. Moreover, information needs regarding tables also contain more entities. This highlights the potentially greater importance of entities in the search of tables than in text. In NQ-TABLES, there is a higher proportion of queries and tables containing entities compared to MS MARCO. It indicates that entities could have a wide-ranging impact on table retrieval, beyond specific or limited cases.

\subsection{Effect of Entities in Relevance Matching} \label{match_pattern}
We aim to illustrate the role of entities in table retrieval. A significant number of entities exist in both queries and tables, yet their role in retrieval remains unclear. We attempt to analyze their effects on table retrieval from the perspective of matching. Specifically, we sampled 300 queries from MS MARCO and NQ-TABLES. For each query, we retained its relevant annotated examples along with 3 irrelevant instances randomly selected from the top 20 results of BM25. Then, we utilized entities appearing in queries to perform exact matching at the entity-level and token-level for relevant and irrelevant examples, respectively: 1) \emph{\textbf{Entity-level}}: For an entity that appears in the query, the match rate is assigned as 1 if it is found in the passage or table, and 0 if it is not.
2) \emph{\textbf{Token-level}}: 
An entity found in a query is tokenized into tokens as the first step. The token-level rate indicates how many of these tokens appear in the passage or table. This statistical approach is closely related to the encoding process since PLMs utilize token sequences as input. We use the tokenizer of BERT to convert the recognized entities into tokens. The statistics results of the two different granularity are presented in Table ~\ref{table:match_rate_between_table_text}. 

\begin{table}[htbp!]
\begin{adjustbox}{width=\linewidth}
\begin{tabular}{lrrrrrr} \toprule
& \multicolumn{3}{c}{Entity-level} & \multicolumn{3}{c}{Token-level} \\
\cmidrule(lr){2-4} \cmidrule(lr){5-7}  
                  & Rel      & Irrel      & $\Delta$($\uparrow$)      & Rel      & Irrel      & $\Delta$($\uparrow$)     \\ \midrule
MS MARCO          & 0.7666 &0.6034 &\textit{0.1632} & 0.8589 &0.7244 &\textit{0.1345} \\
NQ-TABLES         & 0.6954 &0.3992 &\textbf{0.2962} & 0.8352 &0.6701 &\textbf{0.1651}  \\ \bottomrule
\end{tabular}
\end{adjustbox}
\caption{\label{table:match_rate_between_table_text}
Granular Analysis of Entity Matching in Relevant and Irrelevant Examples. Rel and Irrel are matching rates for relevant and irrelevant examples, respectively, $\Delta$ is the gap between them.}
\end{table}

We can see that the gap in matching rates between positive and negative examples is larger
in the table retrieval benchmark than in the text retrieval benchmark whether at the entity level or token level. This suggests that entity entities are more effective in distinguishing relevant tables from irrelevant ones and could be utilized to improve the performance of table retrieval.
When the granularity shifts from entities to tokens, the gap in matching rates between relevant and irrelevant examples decreases. This is because transforming entities into tokens fragments the information and removes the contextual integrity of complete entities. Individual tokens are more likely to appear in irrelevant samples, introducing noise and narrowing the disparity in matching rates.

\begin{figure}[htbp!]
    \centering
    \includegraphics[width=0.9\linewidth]{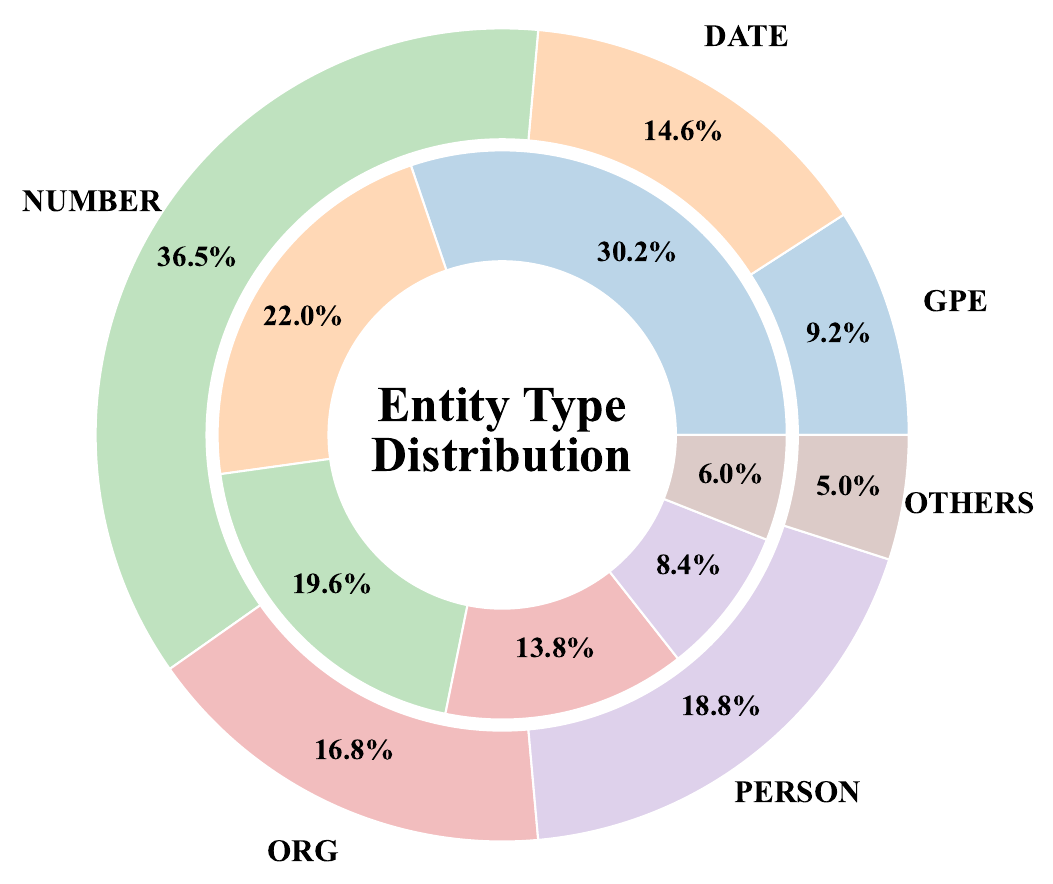}
    \caption{Entity Type Distribution in the Table Retrieval Benchmark. The inner ring visualizes the statistics for queries, whereas the outer ring represents those for tables. Entities of the same type are represented by the same color.
    }
    \label{fig:entity_statistics}
\end{figure}

\subsection{Distribution of Entity Type in Tables} \label{entity_type_distri}
Not all entity types need to be considered in table retrieval, focusing on specific types is sufficient. The table content consists mainly of phrases and words, which include a large number of entities. Due to the table structure, entities of the same type are naturally grouped, such as entities of similar type appearing in the same column. A limited number of entity types often appear in large quantities within a table. Using NQ-TABLES for analysis, we examine the distribution of various entity types within queries and tables. Specifically, we analyze the proportion of each entity type relative to the total number of identified entities. The statistical results are shown in Figure~\ref{fig:entity_statistics}. We can see from this that several entity types, such as number, date, location, and organization, account for a significant proportion of all identified entities. The top 5 most frequent entity types make up more than 80\% of all entities appearing in both queries and tables. Moreover, the entity types with a high proportion are similar between queries and tables. This suggests that we can focus solely on the major entity types in queries and tables in table retrieval.

\section{Entity-Enhanced Training Framework (EE) for Table Retrieval}
\begin{figure*}[htbp!]
    \centering
    \includegraphics[width = \linewidth]{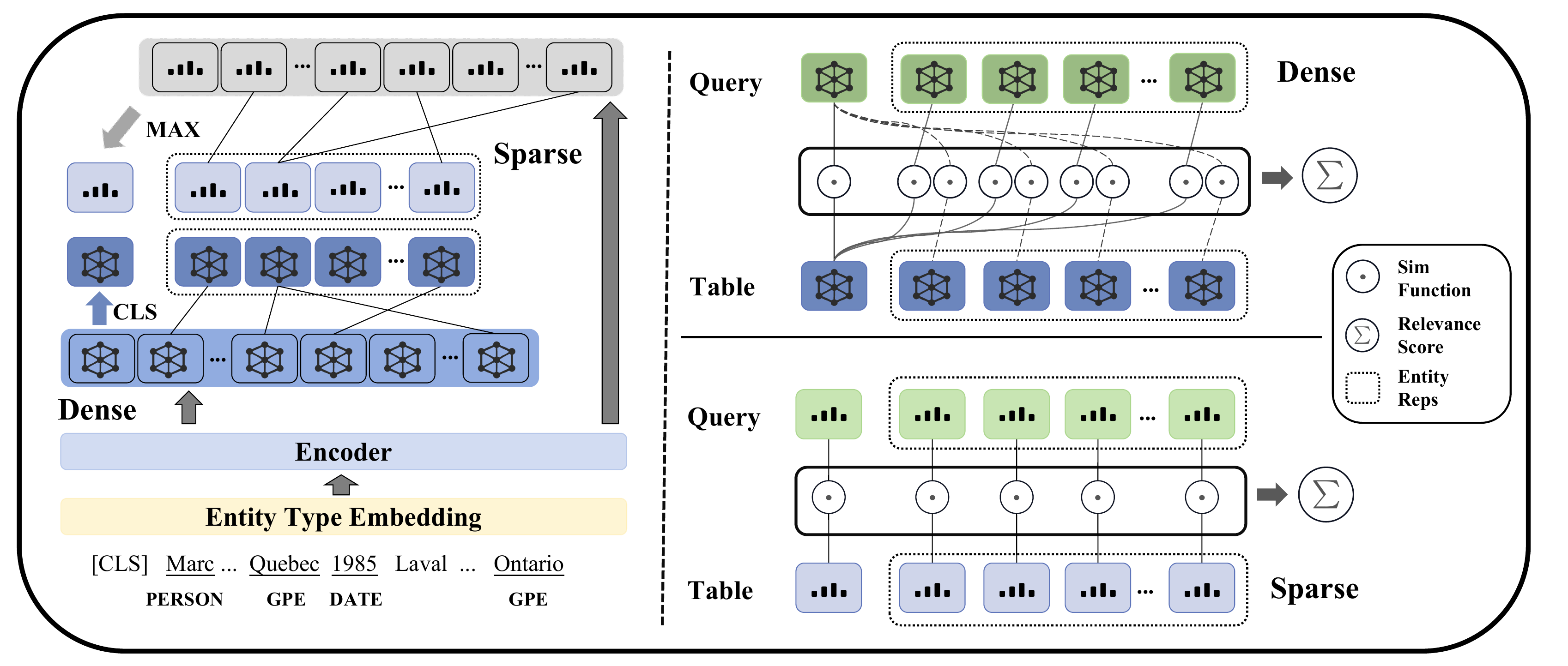}
    \caption{Illustration of Entity-Enhanced Training Framework. Left is the process of encoding and we use the input of the table as an example. The right is the interaction based on entity representations.}
    \label{fig:entity_enhanced_framework}
\end{figure*}

In this section, we describe our proposed entity-enhanced training framework. Through the statistical analysis in Section~\ref{sec:prelim}, we identify the potential role of entities in table retrieval and these insights motivate us to integrate entity information into existing table retrievers to enhance their performance. Table retrievers based on pre-trained language models (PLMs) encode the query and the serialized table, transforming it from a sequence of tokens to a sequence of vectors, a process that can be described as follows:
\begin{equation}
\centering
\begin{aligned}
\mathbf{H}_q = \text{Encoder}(q), \,\mathbf{H}_t = \text{Encoder}(t).
\end{aligned} 
\end{equation}
Different retrievers, including dense and sparse models, are developed by mapping and aggregating the outputs $\mathbf{H}_q$ and $\mathbf{H}_t$. We design an entity-enhanced training framework that can be flexibly adapted to existing table retrievers to improve performance. The framework is shown in Figure~\ref{fig:entity_enhanced_framework}. In the remainder of this section, we will elaborate on the specific design of the framework. 

\subsection{Entity Type Embedding}
The statistical analysis in Section ~\ref{entity_type_distri} reveals that entities frequently appearing in queries and tables are concentrated in a few types, and these types are similar between queries and tables. As a result, it is sufficient to focus on certain types of entities between queries and tables, rather than considering all entities. We retain the top 10 entity types with the highest percentages. Entities in queries and tables are defined as:
\begin{equation}
\centering
\begin{aligned}
&Ent_q = \{(e_i, c_i) \mid c_i \in \mathcal{C},\, i \in \{1, \dots, M\} \}, \\
&Ent_t = \{(e_j, c_j) \mid c_j \in \mathcal{C},\, j \in \{1, \dots, N\}\},
\end{aligned} 
\end{equation} 
where $e_i$ and $e_j$ denote the entities identified in the query and the table, respectively, $c_i$ and $c_j$ represent their entity type (e.g., ‘Person’, ‘Organization’, ‘Location’), M and N correspond to the number of entities in the query and table, respectively. The $\mathcal{C}$ refers to the set of entity types, with $|\mathcal{C}| = 10$.

Some text retrievers incorporate entity information from external knowledge bases during encoding. While they can retain information about entities, their limited coverage makes them impractical for tables, which contain numerous entities. We improve the performance of table retrieval by introducing entity-type information instead of introducing an external knowledge base. Specifically, we propose an entity-type embedding mechanism to inject entity-type information into existing retrievers. This design ensures broader coverage of entities and offers enhanced flexibility for integration with various table retrieval models. After recognizing the entities in both queries and tables, we adopt an adaptive strategy, in which the entity-type embedding is combined with the original input embedding and fed into the retriever. This process can be formalized as:
\begin{equation}    
\centering
\begin{aligned}
&\mathbf{E}_{in} = \text{Emb}_{input}(t),\, \mathbf{E}_{en} = \text{Emb}_{entity}(c),\\
&\mathbf{Gate} = \text{sigmoid}([\mathbf{E}_{in}, \mathbf{E}_{en}]\mathbf{W}_g +\mathbf{b}),\\
&\mathbf{E}_{all} =\mathbf{E}_{in} + \mathbf{Gate}\odot \mathbf{E}_{en},
\end{aligned}
\end{equation} where Emb$_{input}$($\cdot$) converts the input into a sequence of embeddings $\mathbf{E}_{in}$, while Emb$_{entity}$($\cdot$) maps the entity types corresponding to each token in $t$ to a sequence of embeddings $\mathbf{E}_{en}$. For tokens without an associated entity type, a zero-valued entity type embedding is assigned. Each position in $\mathbf{E}_{en}$ is aligned with the corresponding position in $\mathbf{E}_{in}$. We employ a gating mechanism to adaptively integrate the entity type embedding and the token embedding. We use the concatenation of $\mathbf{E}_{in}$ and $\mathbf{E}_{en}$ as input. $\mathbf{W}_g$ and $b$ are the parameters of the linear layer, $\odot$ denotes the element-wise product, and sigmoid($\cdot$) is used to adjust the information about the entity type that flows into the final representation. The entity-enhanced representation $\mathbf{E}_{all}$ replaces the original input representation $\mathbf{E}_{in}$ as input to the PLMs. The encoding process is the same for queries and tables.

\subsection{Interaction for Entity Representation}
The statistics in Section~\ref{match_pattern} demonstrate the important role of entities in distinguishing between relevant and irrelevant tables in different matching ways. To utilize entities during matching, we propose an interaction paradigm based on entity representations, which is flexible and can be integrated with existing table retrievers. 
Depending on the representation used, retrievers can be mainly categorized into dense and sparse models. We illustrate how the interaction approach based on entity representations is designed for the two different types of retrievers, respectively.

\subsubsection{\textbf{Based on Dense Retrievers}}
Dense retrievers use dense vector representations to represent queries and passages. A typical dense text retriever, BIBERT, uses BERT~\cite{devlin2019bert} as a dual encoder to convert queries and tables to dense vectors. The dense retriever in table retrieval follows the same paradigm, and the computation of relevance scores is formalized as:
\begin{equation}
\begin{aligned}
&\mathbf{q}_{ds} = \mathbf{H}_q\text{[CLS]}, \, \mathbf{t}_{ds} = \mathbf{H}_t\text{[CLS]}, \\
&\text{score}_{ds}(q,t) = \text{sim}(\,\mathbf{q}_{ds}, \,\mathbf{t}_{ds}\,),
\end{aligned}
\end{equation} 
where, $q$ represents a query and $t$ represents a table. The representations $\mathbf{q}_{ds}$ and $\mathbf{t}_{ds}$ are the hidden states corresponding to the [CLS] token in the query and table. Using similarity functions $\text{sim}(\cdot)$, such as the inner product or cosine similarity, we compute the relevance score $\text{score}_{ds}(q,t)$ between the query and candidate tables. 

The dense retriever takes the input to the latent space $\mathbb{R}^h$, where $h$ is the dimension of the hidden states generated by the dense retriever. To enrich the entity type information in $\mathbf{q}_{ds}$ and $\mathbf{t}_{ds}$, We design entity-based interactions for $\mathbf{q}_{ds}$ and $\mathbf{t}_{ds}$, respectively. First, we construct the entity representations for queries and tables:
\begin{equation}
\centering
\begin{aligned}
&\mathbf{Entity}_{ds}^{q}[k] = \text{mean}(\mathbf{H}_q[e_i^{start}],...,\mathbf{H}_q[e_i^{end}]),\, \forall c_i = C[k],  \\
&\mathbf{Entity}_{ds}^{t}[k] = \text{mean}(\mathbf{H}_t[e_j^{start}],...,\mathbf{H}_t[e_j^{end}]),\, \forall c_j = C[k],
\end{aligned} 
\end{equation}
where $\mathbf{H}_q$ and $\mathbf{H}_t$ are the corresponding outputs of the query and table by the dense retriever, $e_{i}^{start}$ and $e_{j}^{end}$ represent the start and end indices of the entity in queries, respectively, with $e_j$ specifically referring to tables. We aggregate entities of the same type to streamline the entity representations, $\mathbf{Entity}_{ds}^{q} \in \mathbb{R}^{K \times h}$ and $\mathbf{Entity}_{ds}^{t}$ $\in \mathbb{R}^{K \times h}$, where $K$ denotes the number of entity types. 

Depending on the representation of different entity types in queries and tables, we propose an asymmetric interaction mechanism for dense retrievers. Dense retrievers map the input into vectors in the latent space, making it challenging to quantify the information embedded within these vectors. We aim to optimize the dense representation from the perspective of entity interaction, with an emphasis on highlighting the entity information it contains. Entity-based interactions for dense retrievers are composed of two parts: the interaction between the query representation and the table entity representation, and the interaction between the table representation and the query entity representation. The former emphasizes potentially useful entity information within a query, resembling the concept of query expansion. The latter focuses on entities during information compression, aiming to preserve them as much as possible in the final table representation. These interactions are described as follows:
\begin{equation}
\centering
\begin{aligned}
&\text{score}_{ds}^{q_{e}}(q,t) = \sum_{k =1}^{|\mathcal{C}|} \text{sim}(\mathbf{q}_{ds}, \mathbf{Entity}_{ds}^{t}[k]),\\
&\text{score}_{ds}^{t_{e}}(q,t) = \sum_{k =1}^{|\mathcal{C}|} \text{sim}(\mathbf{t}_{ds}, \mathbf{Entity}_{ds}^{q}[k]),\\
&\text{score}_{ds}^{train}(q,t) = \text{score}_{ds}(q,t) + \lambda_{ds}^{q_{e}}* \text{score}_{ds}^{q_{e}}(q,t) \\
& \hspace{1.8cm} + \lambda_{ds}^{t_{e}}*\text{score}_{ds}^{t_{e}}(q,t),
\end{aligned} 
\end{equation} 
where $\text{score}_{ds}^{q_{e}}(q,t)$ denotes the relevance score between the query representation and the table entity representations, calculated as the sum of relevance scores between the query representation and each entity type representation. Similarly, $\text{score}_{ds}^{t_{e}}(q,t)$ represents the relevance score between the query entity representation and the table representation, $\lambda_{ds}^{q_{e}}$ and $\lambda_{ds}^{t_{e}}$ are the corresponding weights during training.

\subsubsection{\textbf{Based on Sparse Retrievers}}
Sparse retrievers use sparse vector representations to represent queries and passages, which means that most elements in representations are zero, and only a few occurrences of words or features are represented. SPLADE~\citep{formal2021splade} transforms the input into a distribution in the vocabulary space, typically generated by mapping the hidden states from the output of the PLMs through a projection layer, the sparse representation can be obtained as follows:
\begin{equation}
\begin{aligned}
\label{eq:sparse_emb}
&\mathbf{S}_{q} = \text{transform}(\mathbf{H}_{q}), \,\mathbf{S}_{t} = \text{transform}(\mathbf{H}_{t}), \\
&\text{transform}(\mathbf{X}) = \mathbf{X}\mathbf{A}^\top + \mathbf{b},
\end{aligned}
\end{equation}
where $\mathbf{H_q}\in \mathbb{R}^{|q| \times h}$ and $\mathbf{H_t}\in \mathbb{R}^{|t| \times h}$ represent the sequence of vectors of dimension $h$ yielded by PLMs. 
$\mathbf{S}_{q}\in \mathbb{R}^{|q| \times |V|}$ and $\mathbf{S}_{t}\in \mathbb{R}^{|t| \times |V|}$ correspond to the logits of query $q$ and table $t$ from the output of the transform ($\cdot$) respectively. $\mathbf{A} \in \mathbb{R}^{|V| \times h}$ is the parameters and b is the bias term of a linear layer. 
Sparse representations of $q$ and $t$, i.e., $\mathbf{q}_{sps}$ and $\mathbf{t}_{sps}$, are constructed by applying pooling (mean or max) over the entire sequence:
\begin{equation}    
\begin{aligned}
\label{eq:sparse_qd}
&\mathbf{q}_{sps} = \text{Pooling}\,(\text{ReLU}(\mathbf{S}_{q})),\\
&\mathbf{t}_{sps} = \text{Pooling}\,(\text{ReLU}(\mathbf{S}_{t})),\\
&\text{score}_{sps}(q,t) = \text{sim}(\,\mathbf{q}_{sps}, \,\mathbf{t}_{sps}\,),
\end{aligned}
\end{equation}
where $\text{score}_{sps}(q,t)$ is the relevance score between the query $q$ and table $t$ based on the sparse representation. Similar to dense retrievers, we employ the mean pooling to construct sparse representations for different entity types from $\mathbf{S}_{q}$ and $\mathbf{S}_{t}$:
\begin{equation}
\centering
\begin{aligned}
&\mathbf{Entity}_{sps}^{q}[k] = \text{mean}(\mathbf{S}_q[e_i^{start}],...,\mathbf{S}_q[e_i^{end}]),\, \forall c_i = C[k],\\
&\mathbf{Entity}_{sps}^{t}[k] = \text{mean}(\mathbf{S}_t[e_j^{start}],...,\mathbf{S}_t[e_j^{end}]),\, \forall c_j = C[k],
\end{aligned}
\end{equation}
where $\mathbf{Entity}_{sps}^{q} \in \mathbb{R}^{K \times |V|}$ and $\mathbf{Entity}_{sps}^{t} \in \mathbb{R}^{K \times |V|}$ denote the sparse representations of different entity types in the query and the table, $K$ is the number of entity types. 

SPLADE maps the input to become a distribution over the vocabulary space. We can visually examine which tokens are retained in the sparse representation. This indicates that we only need to enhance the interaction between tokens of the same type of entity in the query and the table, thus highlighting the entity information in the final sparse representation:
\begin{equation}
\begin{aligned}
\text{score}_{sps}^{e}(q,t) &=  \sum_{k =1}^{|\mathcal{C}|} \text{sim}(\,\mathbf{Entity}_{sps}^{q}[k], \, \mathbf{Entity}_{sps}^{t}[k]\,),\\
\text{score}_{sps}^{train}(q,t) &= \text{score}_{sps}(q,t) + \lambda_{sps}^{e}*\text{score}_{sps}^{e}(q,t),
\end{aligned}
\end{equation} 
where $\text{score}_{sps}^{e}(q,t) $ is defined as the sum of relevance scores between entity representations of the same type and $\lambda_{sps}^{e}$ is the weight. 

\subsection{Training and Inference}
\subsubsection{\textbf{Training Objective.}} During the training process, we incorporate entity-type embeddings and design an entity-enhanced interaction paradigm based on the outputs of different types of retrievers. The training process aims to minimize the InfoNCE~\citep{he2020momentumcontrastunsupervisedvisual} loss. For a query $q_i$ in a batch, we pair the positive table $t_i^{+}$ with a set of random negative tables (e.g., positive tables from other queries in the batch, ${t_{j}^+}$ for query $q_j$), denoted as ${t_{i,j}^-}$). The loss is computed as:
\begin{equation}
\centering
\begin{aligned}
\ell &= -log \frac{e^{s(q_i,t_i^+)}}{e^{s(q_i,t_i^+)}+\sum_je^{s(q_i,t_{i,j}^-)}},\\
s(q,t) &= \begin{cases} 
\text{score}_{ds}^{train}(q,t) & \text{for dense retrievers}, \\
\text{score}_{sps}^{train}(q,t) & \text{for sparse retrievers.}
\end{cases}
\end{aligned}
\end{equation} 

\subsubsection{\textbf{Inference Score.}} Unlike the training phase, which involves multiple relevance scores, we use $\text{score}_{ds}(q,t)$ or $\text{score}_{sps}(q,t)$ as the relevance score between the query and the table during inference. Our proposed entity-enhanced training framework aims to emphasize entity-related information in the representation rather than introducing complex interactions. Only the embedding of entity types was retained in the inference phase.

\subsubsection{\textbf{Differences from Multi-Vector Retrievers.}} Although our proposed entity-enhanced training framework is similar to multi-vector retrievers in the training phase, they are fundamentally distinct. Multi-vector retrievers are designed to facilitate fine-grained interaction of queries and tables by generating multiple vectors. Our framework, however, aims to refine representations of existing retrievers. Although their training processes are similar, their inference procedures differ significantly. Multi-vector retrievers use the same interaction mechanism for relevance score computation during inference as in training, resulting in inference costs that are several times higher than those of single-vector retrievers. By contrast, our framework adopts the same relevance score computation method as single-vector retrievers, resulting in lower costs and better performance than multi-vector retrievers.
\section{Experimental Setup}
\subsection{Datasets}
We conduct experiments on two standard table retrieval benchmarks:
\begin{itemize}[leftmargin=*]
    \item \textbf{NQ-TABLES~\citep{herzig2021open}} is a subset of the Natural Questions (NQ)~\citep{kwiatkowski-etal-2019-natural}, a QA dataset based on Wikipedia. NQ is collected from real web pages and search logs, with relevant annotations and answers corresponding to the queries. NQ-TABLES is the data associated with the table. It can be used as a benchmark for table retrieval and table QA.
    \item \textbf{OTT-QA~\citep{kostić2021multimodal}} is an open-domain, multi-hop QA dataset that includes both text and tables from Wikipedia. Derived from a closed-domain QA dataset, it simplifies the relevance annotation by decontextualizing the question and treating the reference table as a correlation table, overlooking other relevant examples. We use the subset of table-related data for our evaluation.
\end{itemize}
The statistics of benchmarks are shown in Table~\ref{table:benchmark_static}.
\begin{table}[htbp!]
\centering
\resizebox{\linewidth}{!}{
\begin{tabular}{llrrrr} \toprule
                                             &                              & \multicolumn{2}{c}{\textbf{NQ-TABLES}} & \multicolumn{2}{c}{\textbf{OTT-QA}} \\
\cmidrule(lr){3-4} \cmidrule(lr){5-6} 
                                             &                                        & Train          & Test         & Train        & Test        \\ \midrule
\multicolumn{1}{l|}{\multirow{2}{*}{Query}}  & \multicolumn{1}{l|}{Count}             & 9,594           & 919          & 41,469        & 2,214        \\
\multicolumn{1}{l|}{}                        & \multicolumn{1}{l|}{Avg. \# Words.}  & 8.94           & 8.90        & 21.79        & 22.82      \\ \midrule
\multicolumn{1}{l|}{\multirow{3}{*}{Table}} & \multicolumn{1}{l|}{Count}             & 169,898         & 169,898       & 419,183       & 419,183      \\
\multicolumn{1}{l|}{}                        & \multicolumn{1}{l|}{Avg. \# Row.}      & 10.70       & 10.70    & 12.90   & 12.90   \\ 
\multicolumn{1}{l|}{}                        & \multicolumn{1}{l|}{Avg. \# Col.}      & 6.10       & 6.10    & 4.80   & 4.80   \\ \midrule
\multicolumn{2}{l|}{\# Golden Tables per Query}                                                 & 1.00            &1.05          & 1.00         & 1.00         \\ \bottomrule
\end{tabular}}
\caption{\label{table:benchmark_static}
Statistics of the Table Retrieval Benchmarks. 
}
\end{table}

\subsection{Baselines}
We use different backbones of retriever to demonstrate the effectiveness and generalizability of our entity-enhanced training framework, these retrievers can be divided into two categories: sparse and dense. \\
\textbf{Sparse retrievers:}
\begin{itemize}[leftmargin=*]
    \item \textbf{BM25~\citep{robertson1994some}} is a widely used sparse retrieval method that estimates the relevance of documents to a user query based on bag-of-words representations and exact term matching.
    \item \textbf{SPLADE~\citep{formal2021splade}} learned sparse retriever based on PLMs such as BERT. It maps a query or table to a vector of vocabulary size, where each dimension represents the probability of a specific term.
\end{itemize}
\textbf{Dense retrievers:}
\begin{itemize}[leftmargin=*]
    \item \textbf{BIBERT~\citep{bibert-pretrain}} is a standard dense retriever based on BERT, the relevance score between a query and a table is estimated using the hidden state of [CLS] from the output of BERT.
    \item \textbf{TAPAS~\citep{Herzig_2020}} utilizes distinct embeddings, such as row and column embeddings, to represent table structure. It is pre-trained on a large corpus of tabular data and is a universal table encoder widely used in table-related tasks.
    \item \textbf{DTR~\citep{herzig2021open}} uses Inverse Cloze Task (ICT) for pre-training based on TAPAS. It also down-projects the final query and table representations to optimize effectiveness and efficiency.
    \item \textbf{SSDR$_{im}$~\citep{jin2023enhancing}} is a multi-vector retriever that converts queries and tables into multiple vectors, enhancing the single-vector retriever’s capability to capture detailed information from both queries and tables. The relevance score is an aggregation of the scores obtained by one-to-one matching between two vector lists.
\end{itemize}
We conduct a comparative analysis of four retrievers (SPLADE, BIBERT, TAPAS, and DTR) to demonstrate the effectiveness of our proposed entity-enhanced training framework. These retrievers represent different optimization directions for table retrieval and have been widely used in real-world scenarios. There are also many optional retrievers such as BGE~\citep{chen2024bgem3embeddingmultilingualmultifunctionality}, LLM2Vec~\citep{behnamghader2024llm2veclargelanguagemodels}, etc. which use more data to achieve significant retrieval performance improvements at larger models. However, our entity-enhanced training framework is orthogonal to the optimization of these retrievers. 

\subsection{Evaluation Metrics}
For retrieval evaluation, we use recall (Recall) and normalized discounted cumulative gain (NDCG). Since the retrieved tables will be used in downstream tasks such as table comprehension and question answering, we use 50 as the maximum cutoff, following previous studies~\citep{Herzig_2020,jin2023enhancing} that also evaluate the top 50 results. Specifically, we report Recall@1, Recall@10, and Recall@50 to measure the number of relevant tables retrieved, and NDCG@3 and NDCG@5 to assess whether the most relevant tables are ranked in the top positions. We also demonstrate the effectiveness of our proposed entity-enhancement framework by evaluating the performance of end-to-end QA. In addition to relevance annotations, NQ-Tables also provide answers that are factual, concise, and precise. We use accuracy as a metric to analyze the performance of different retrievers within the RAG system.

\subsection{Implementation Details}
We initialize the baselines using publicly available checkpoints and compare their performance with and without our proposed entity-enhanced training framework. To ensure a fair comparison, all PLMs are trained with a batch size of 144 and a learning rate of $1e^{-5}$. We compare the performance of different baselines after 50 training epochs.
Due to the large vocabulary space of SPLADE, we introduce the $FLOPS$ regularizer~\citep{paria2020minimizing} that constrains the number of non-zero terms in its representation. We use spaCy to recognize entities that appear in queries and tables. 
During training, $\lambda_{ds}^{q_{e}}$ and $\lambda_{ds}^{t_{e}}$ are used to control the weight of entity matching scores in the relevance scores of dense retrievers, while $\lambda_{sps}^{e}$ is specifically designed for sparse retrievers. Both of them are determined using a grid search. For inference, Faiss~\citep{2022ascl.soft10024J} is employed for approximate nearest neighbor searches in dense retrieval, while inverted indexes are utilized for sparse retrieval.

\begin{table*}[htbp!]
\centering
\resizebox{\linewidth}{!}{
\begin{tabular}{lllllllllll} \toprule
\renewcommand{\arraystretch}{1.1}
             & \multicolumn{5}{c}{\textbf{NQ-TABLES}}            & \multicolumn{5}{c}{\textbf{OTT-QA}}              \\
\cmidrule(lr){2-6} \cmidrule(lr){7-11}                                                       
             & NDCG@3 &NDCG@5 & Recall@1 & Recall@10 & Recall@50 & NDCG@3 & NDCG@5 & Recall@1 & Recall@10 & Recall@50 \\ \hline
BM25         & 23.59   &25.52    & 17.95  & 37.05     & 52.61     & 32.65   &35.09  & 23.98   & 51.94     & 69.11   \\
DTR         &44.70   &49.13   &31.72  &74.63  &90.99  &54.62  &57.23 & 42.14   &75.93     &87.58      \\
SPLADE       & 54.50   &58.46    & 38.43  & 83.13     & 95.25     & \underline{74.65}   &\underline{76.31}  &\underline{64.45}   & \underline{89.39}     & \textbf{95.89}      \\  
TAPAS       &56.93  &61.08       &43.91  & 83.55     & 95.21  & 68.09   &70.44  & 56.05 &86.72 &94.49     \\
BIBERT         & 57.53   &61.73    & 44.49  & 84.87    & 94.16     & 68.37  &70.28    &57.05   & 86.45     & 94.26      \\
SSDR$_{im}$    &\underline{58.50}  &\underline{61.89}    &\underline{45.14} & 84.71 &\textbf{95.77}    &67.53   &69.81  &56.96  &86.22     &93.95\\
\midrule

EE-DTR  & 47.37$^{\dagger}$   &51.47$^{\dagger}$  &33.73$^{\dagger}$  &75.44  & 91.07     &55.60$^{\dagger}$   &58.00$^{\dagger}$     &43.77$^{\dagger}$ &76.69   &86.95 \\
EE-SPLADE       &57.26$^{\dagger}$   &60.70$^{\dagger}$     &42.29$^{\dagger}$   & 84.37     & \underline{95.38}     &\textbf{74.98}  &\textbf{76.40}     &\textbf{64.50}   &\textbf{89.52}     &\underline{95.57}     \\
EE-TAPAS & 59.74$^{\dagger}$   &63.63$^{\dagger}$  &45.90$^{\dagger}$  &\textbf{85.56}    & 94.74       &69.83$^{\dagger}$  &71.76$^{\dagger}$    &58.81$^{\dagger}$   &86.95     &94.40 \\
EE-BIBERT & \textbf{60.43}$^{\dagger,\star}$   &\textbf{64.46}$^{\dagger,\star}$    & \textbf{47.53}$^{\dagger,\star}$   & \underline{85.35}     &94.85     &68.77  &70.81  &57.23   &86.72    &94.08 \\

\bottomrule
\end{tabular}}
\caption{\label{table:main_result}
Overall table retrieval performance. Bold and \underline{underline} indicate the optimal performance and suboptimal performance respectively. `$\dagger$' indicates statistically significant differences (p<0.05) between the entity-enhanced retriever and the vanilla retriever. '$\star$' represents statistically significant differences (p<0.05) compared to best baseline.}
\end{table*}

\section{Overall Performance}
We compare the performance of different baselines with and without our training framework, respectively. The complete results are presented in Table~\ref{table:main_result}. As shown in the table, both dense and sparse retrievers demonstrate improved performance under our proposed entity-enhanced training framework. This illustrates the effectiveness and generalizability of our proposed framework. 
Our proposed framework achieves substantial improvements in NDCG@3 and NDCG@5, while maintaining slightly improved performance in recall metrics for high cutoffs. These results suggest that our framework prioritizes the improvement of retrieval quality. Our training framework focuses solely on the entities present in the query or table without introducing any additional information. This may indicate a preference for relying on intrinsic entity matching. 
If the goal is to improve both NDCG and Recall of different retrievers through entities, entities that are absent from queries and tables could be considered in the encoding process.

There are also some interesting observations through horizontal and vertical comparisons: 1) In general, dense retrievers outperform sparse retrievers in text retrieval due to their ability to perform semantic matching rather than term matching. This is not always the case in table retrieval scenarios. On the NQ-TABLES dataset, BIBERT outperforms SPLADE, whereas the opposite result is observed on the OTT-QA dataset. This is caused by queries with different characteristics. Queries in OTT-QA are longer and contain a lot of detailed information that is indispensable in relevance matching. This makes term-matching exceptionally important. Even the multi-vector dense retriever $SSDR_{im}$ does not perform well enough in capturing details. SPLADE has the best performance among all baselines on OTT-QA. This is also the reason why our proposed entity-enhanced training framework has fewer improvements on OTT-QA.
2) Despite the differences in the calculation of relevance scores between training and inference, the retrieval performance still shows improvement. This shows that our proposed entity enhancement training framework can adjust the original query and table representations rather than merely combining relevance scores from different perspectives.
\section{Further Analyses}
We conduct experiments with different variants of the framework to evaluate whether the components of our proposed framework have positive effects. There are three aspects we aim to investigate: 1) \emph{\textbf{What is the contribution of each component in our proposed framework to the performance of table retrieval?}} 2) \emph{\textbf{How do different inference patterns influence the performance of table retrieval?}} 3) \emph{\textbf{How does the weight of the relevance score based on entities impact retrieval performance?}}

\textbf{Contribution of Different Components.}
Our proposed framework incorporates various components, including entity type embedding and late interactions between entities in queries and tables. To evaluate the individual contributions of these components, we removed each component from the original training framework one by one and retrained the retriever. The performance of different variants is shown in Table~\ref{tab:results-ablation_components}. 
It demonstrates that each component independently contributes to improving retrieval performance. The results of BIBERT and SPLADE further highlight the importance of interactions between entities. The interaction for dense retrievers is asymmetric, and the results of BIBERT indicate that the interaction between the table representation and the entities in the query is more important. PLMs are proficient in compressing. This interaction aligns with the compression process. When entity type embedding is excluded from our training framework, both EE-BIBERT and EE-SPLADE show a decline in performance, underscoring the insufficiency of entity knowledge in PLMs.

\begin{table}[htbp!]
\centering
\begin{adjustbox}{width=\linewidth}
\renewcommand{\arraystretch}{1.2}
\begin{tabular}{llll} \toprule
                      & NDCG@3   & NDCG@5 & Recall@10 \\ \midrule
EE-BIBERT             &\textbf{60.43}$^{\dagger}$        &\textbf{64.46}$^{\dagger}$          &\textbf{85.35}   \\
\, w/o score$_{ds}^{q_{e}}$         &\underline{59.91}$^{\dagger}$        &\underline{63.83}$^{\dagger}$           &84.58   \\
\, w/o score$_{ds}^{t_{e}}$         &59.10       &62.85           &84.75   \\
\, w/o Emb$_{entity}$(q\&t)       &58.46        &62.60           &84.60   \\
BIBERT                               &58.12        &62.37           &84.74   \\ \midrule 

EE-SPLADE            &\textbf{57.26}$^{\dagger}$         &\textbf{60.70}$^{\dagger}$  &\textbf{84.37}   \\
\, w/o score$_{sps}^{e}$       &54.45        &58.81           &83.88   \\
\, w/o Emb$_{entity}$(q\&t)        &\underline{56.59}$^{\dagger}$      &\underline{60.30}$^{\dagger}$        &\underline{83.32}   \\
SPLADE                              &54.50        &58.46           &83.13   \\ \bottomrule 

\end{tabular}
\end{adjustbox}
\caption{Ablation for Different Components. $\dagger$ denotes statistically significant differences compared to BIBERT or SPLADE, respectively.}
\label{tab:results-ablation_components}
\end{table}

\textbf{Analysis of Various Inference Patterns.}
During training, our proposed framework incorporates entity-type embedding and interaction mechanisms based on entity types. However, only the entity type embedding is kept in the inference stage. Although tables can be encoded offline, queries are processed online. Entities appearing in a query need to be identified before encoding, which may introduce additional latency. We attempt to remove the entity type embedding of queries and tables during inference for the retrievers trained with our framework. The corresponding performance is presented in Table~\ref{tab:result-ablation_inference}. It indicates that entity type embedding has a limited effect on retriever performance in the inference phase. Although removing the entity type embeddings from EE-BIBERT and EE-SPLADE results in reduced performance, the performance remains higher than the BIBERT and SPLADE. We can improve retriever efficiency by omitting the entity type embedding. 
\begin{table}[htbp!]
\centering
\begin{adjustbox}{width=\linewidth}
\renewcommand{\arraystretch}{1.1}
\begin{tabular}{llll} \toprule
                      & NDCG@3 & NDCG@5 & Recall@10 \\ \midrule
EE-BIBERT                &\textbf{60.43}$^{\dagger}$  & \textbf{64.46}$^{\dagger}$       & \textbf{85.35}       \\
\, w/o Emb$_{entity}$(q)       &\underline{60.42}$^{\dagger}$           & \underline{64.02}$^{\dagger}$          & 84.87   \\
\, w/o Emb$_{entity}$(q\&t)       &60.27$^{\dagger}$           & 63.79$^{\dagger}$          & 84.75   \\ 
BIBERT                            &58.12           & 62.37     & 84.87   \\ \midrule
EE-SPLADE                 &\textbf{57.26}$^{\dagger}$  &\textbf{60.70}$^{\dagger}$           &\textbf{84.37}       \\
\, w/o Emb$_{entity}$(q)       &56.68           & 60.05          & \underline{84.26}   \\
\, w/o Emb$_{entity}$(q\&t)    &\underline{56.86}           & \underline{60.11}          & 83.71   \\ 
SPLADE                            &54.50           &58.46           &83.13    \\ \bottomrule
\end{tabular}
\end{adjustbox}
\caption{Evaluation of Different Inference Patterns. $\dagger$ indicates statistically significant compared to backbone models.}
\label{tab:result-ablation_inference}
\end{table}

\begin{figure}[htbp!]
\includegraphics[width=0.9\linewidth]{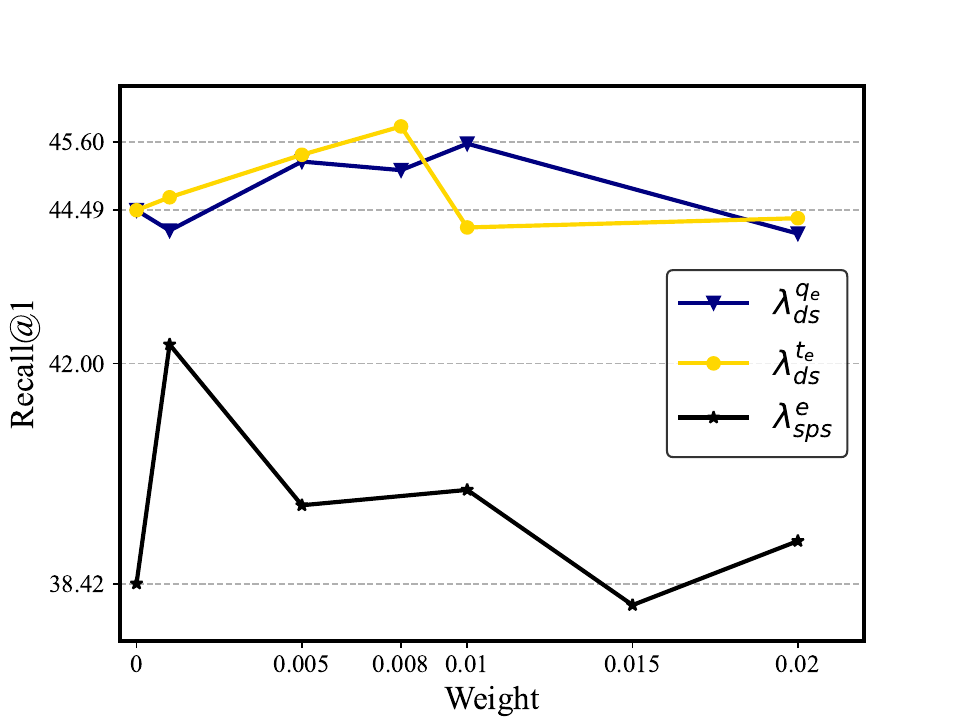}
\caption{Effect of Entity Matching Weights on BIBERT and SPLADE.}
\label{fig:hyper-params}
\end{figure}

\textbf{Impact of Weights of Entity Matching Scores.}
For dense retrievers, we propose asymmetric interactions based on entity representations: $\text{score}_{ds}^{q_{e}}$ and $\text{score}_{ds}^{t_{e}}$, which are designed separately for queries and tables. We introduce weights $\lambda_{ds}^{q_{e}}$ and $\lambda_{ds}^{t_{e}}$ to adjust their contribution in the calculation of the relevance score during training. We define $\text{score}_{sps}^{e}$ and $\lambda_{sps}^{e}$ for sparse retrievers to serve the comparable role. To analyze the impact of weights on our proposed training framework, we integrate each relevance score individually into the vanilla retriever and compare their performance after training. The performance of Recall@1 is shown in Figure~\ref{fig:hyper-params}. For BIBERT and SPLADE, appropriate weights can also improve its retrieval performance. However, inappropriate weights can degrade the performance of two different types of retrievers. Interestingly, for BIBERT, $\lambda_{ds}^{t_{e}}$ provides a greater performance improvement for BIBERT retrieval than $\lambda_{ds}^{q_{e}}$. This is because $\text{score}_{ds}^{t_{e}}$ is intended to preserve entity information in the process of compressing the table. This is analogous to the function of information compression in PLMs. 

\section{Case Study}
\begin{figure}[htbp!]
  \centering
  \includegraphics[width= \linewidth]{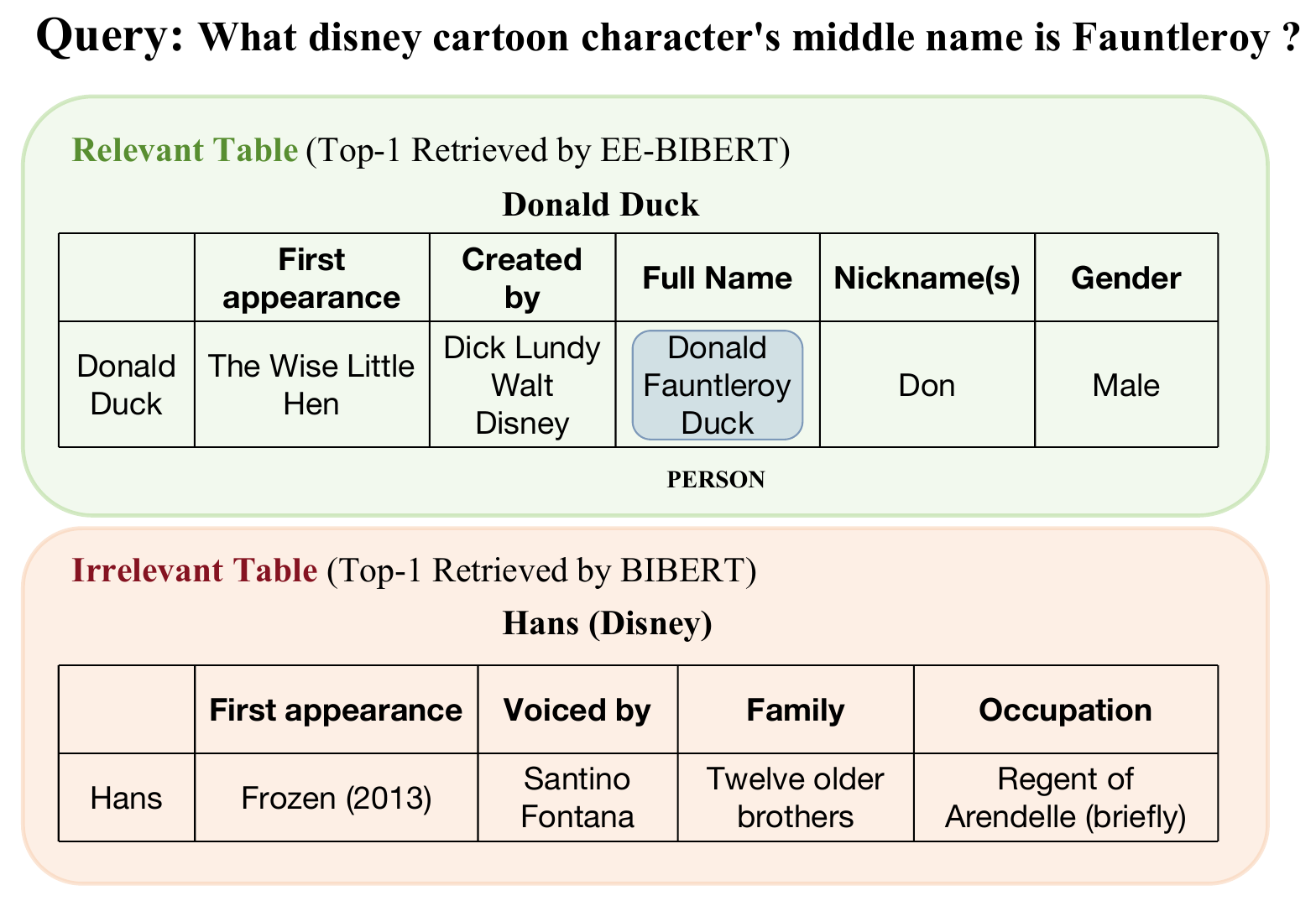}
  \caption{Retrieval Result Comparison between EE-BIBIERT and BIBERT.}
  \label{fig:case_study}
\end{figure}

To illustrate how our proposed training framework enhances retrieval performance, we provide a concrete example in Figure~\ref{fig:case_study}. Below are the tables retrieved by EE-BIBERT and BIBERT that ranked first in the retrieval list. The relevant table does not include ``disney'', while the irrelevant table matches ``disney'' in the title. The title usually has more importance than a specific cell during matching. ``Donald Fauntleroy Duck'' in the cell of the relevant table is identified as the entity of type \textbf{PERSON}. Our proposed entity-enhanced training framework highlights this entity in the table representation. After training, EE-BIBERT becomes more sensitive to ``Fauntleroy'' in the query,  allowing it to retrieve the most relevant table containing ‘Donald Fauntleroy Duck’ at the top position.
\section{Application and Impact in TableQA}
We conducted an end-to-end evaluation of the TableQA to demonstrate the practical value of our entity-enhanced training framework in real-world scenarios. Mistral~\citep{jiang2023mistral7b}, Llama3~\citep{grattafiori2024llama3herdmodels}, and Qwen-2.5~\citep{qwen2025qwen25technicalreport} are used to conduct retrieval-augmented generation in an end-to-end manner. The evaluation of responses is based on the top tables retrieved by different retrievers, and the results are summarized in Table~\ref{table: application}.
\begin{table}[htbp!]
\begin{adjustbox}{width=\linewidth}
\renewcommand{\arraystretch}{0.90}
\begin{tabular}{lcrrr} \toprule
\multirow{2}{*}{Retriever}          & \multirow{2}{*}{LLM} & \multicolumn{3}{c}{Accuracy} \\ \cmidrule(lr){3-5}
                                    &                      & n=1     & n=3     & n=5     \\ \midrule
\multirow{3}{*}{SSDR$_{im}$}        & Mistral-7B              &0.3276   &0.3695   &0.3630    \\
                                    & Llama3-8B               &0.3425   &0.3814   &0.3789    \\
                                    & Qwen2.5-7B             &0.3342   &0.3927   &0.3966          \\ \midrule                                        
\multirow{3}{*}{SPLADE}             & Mistral-7B              &0.2961   &0.3542   &0.3495    \\
                                    & Llama3-8B               &0.3207   &0.3717   &0.3388    \\
                                    & Qwen2.5-7B          &0.3190   &0.3579   &0.3735          \\ \midrule
\multirow{3}{*}{EE-SPLADE}  & Mistral-7B                      &0.3176   &0.3633   &0.3602    \\
                                    & Llama3-8B               &0.3383   &0.3723   &0.3924    \\
                                    & Qwen2.5-7B             &0.3191   &0.3822   &\underline{0.3989} \\ \midrule    
\multirow{3}{*}{BIBERT}             & Mistral-7B              &0.3293   &0.3346   &0.3530    \\
                                    & Llama3-8B               &0.3266   &0.3316   &0.3428    \\
                                    & Qwen2.5-7B             &0.3480   &0.3792   &0.3701          \\ \midrule                        
\multirow{3}{*}{EE-BIBERT} & Mistral-7B              &0.3482   &0.3771   &0.3758    \\
                                    & Llama3-8B              &\textbf{0.3718}   &\underline{0.4007}   &\textbf{0.3994}    \\
                                    & Qwen2.5-7B            &\underline{0.3665}  &\textbf{0.4106}         &\underline{0.3989}           \\ \bottomrule
\end{tabular}
\end{adjustbox}
\caption{\label{table: application} End-to-end QA Performance of NQ-TABLES, n indicates the number of tables retrieved by various retrievers in the context.}
\end{table}
In general, LLMs generate more accurate answers when retrieval performance is improved. However, this is not always the case. Different LLMs demonstrate varying abilities in parsing tables. LLMs may generate incorrect answers even when the relevant table is in context. Increasing the number of tables in the context does not always cause LLMs to generate better answers. Although adding more tables raises the likelihood of including relevant ones, it simultaneously introduces noise from irrelevant tables, emphasizing the importance of high-quality retrieval. Our proposed entity-enhanced training framework improves retrieval quality by leveraging fine-grained entity matching, making it well-suited for RAG systems.

\section{Conclusion and Future Work}
Table retrieval is crucial in accessing vast amounts of information stored within tables.
The table content consists mainly of phrases and words, which include a large number of entities. Whether these entities can be leveraged to enhance retrieval performance has not been thoroughly explored. In this work, we adopt a statistical perspective to analyze and highlight the significant role entities play in table retrieval. At the same time, we design an entity-enhanced training framework that is plug-and-play and flexible enough to integrate with existing table retrievers. Extensive experiments demonstrate the generalizability and effectiveness of our proposed framework.

The table is one of the many formats used to store information. Together with other formats such as HTML, text, and PDF, it constitutes a vast amount of data in the real world. Existing table retrievers are mainly for tables stored as text. How to effectively retrieve tables in other formats such as images, PDFs, etc. has still not been effectively explored. LLMs significantly expand the scope of downstream applications for table retrieval. Although LLM's ability to handle complex tasks has improved dramatically, it still falls short in understanding the content of tables. Optimizing table retrieval and enhancing the ability of LLMs to process tables are long-term objectives for a table-centric RAG system.

\clearpage
\bibliographystyle{ACM-Reference-Format}
\balance
\bibliography{main}
\end{document}